\documentclass[aps,showpacs,preprintnumbers,amsmath,amssymb,superscriptaddress,floatfix,a4paper,nofootinbib,11pt,floatfix,nofootinbib,onecolumn]{revtex4}
\usepackage{graphicx}
\usepackage{bm}
\usepackage{rotating}
\usepackage{array}
\usepackage{xcolor}
\usepackage{amsmath,amssymb}
\usepackage{mathrsfs}
\usepackage{graphicx}
\usepackage{color}
\usepackage{subfigure}
\usepackage{fancyhdr}
\usepackage{multirow}
\usepackage{float}
\usepackage{epsfig}
\usepackage{amsfonts}
\usepackage{bm}
\usepackage{multirow}
\usepackage{nicefrac}

\newcommand{\ba}{\begin{eqnarray}}
\newcommand{\ea}{\end{eqnarray}}

\begin{document}

\title{Refined Swampland conjecture in deformed Starobinsky gravity}

\author{Phongpichit Channuie} 
\email{channuie@gmail.com}
\affiliation{College of Graduate Studies, Walailak University, Thasala, Nakhon Si Thammarat, 80160, Thailand}
\affiliation{School of Science, Walailak University, Thasala, \\Nakhon Si Thammarat, 80160, Thailand}

\date{\today}

\begin{abstract}

In order to validate or invalidate a
large class of low energy effective theories, the Swampland conjecture has attracted significant attention, recently. It can be stated as inequalities on the potential of a scalar field which is conjectured
to satisfy certain constraints. In this work, we discuss the theoretical viability of deformed Starobinky gravity in light of the refined Swampland conjectures. We consider the deformation of the form $f(R)\sim R^{2(1-\alpha)}$ with $\alpha$ being a constant. We then constrain $\alpha$ using the spectral index of curvature perturbation $n_s$ and the tensor-to-scalar
ratio $r$. We demonstrate that the model under consideration is
in strong tension with the refined swampland conjecture. However, regarding our analysis with proper choices of parameters  $a,\,b=1- a$ and $q$, we discover that the model can always satisfy this new refined swampland conjecture. Therefore, the model might be in “landscape” since the “further refining de Sitter swampland conjecture” is satisfied.

\end{abstract}

%\pacs{Valid PACS appear here}

\maketitle

%%%%%%%%%%%%%%%%%%%%%%%%%%%%%%%%%%%%%%%%

\section{Introduction}
In string theory, it has been suggested that a landscape of vacua is vast and consistent quantum theory of gravity is believed to be formulated with consistent low-energy effective field theories (EFTs). Even more recently, the authors of Refs.\cite{Obied:2018sgi,Agrawal:2018own,Ooguri:2018wrx} suggested that the landscape is possibly surrounded by an even more vast neighborhood called \lq\lq swampland \rq\rq where consistent EFTs, which are coupled to gravity, are inconsistent with quantum theory of gravity. To be more concrete, the swampland can be formulated by the set of consistent effective field theories which cannot be completed into any quantum gravity in the high energy regime. Hence, it is desirable for consistent EFTs not to lie in the swamplands. See a comprehensive review on the Swampland \cite{Palti:2019pca}.

In light of recently proposed swampland conjectures which
can be translated to inequalities on the potential for the
scalar field driving inflation. To start our discussion, we first follow a setup given in Ref.\cite{Andriot:2018mav}, and consider a four-dimensional $(4D)$ theory of real scalar field $\phi^{i}$ coupled to gravity, and its dynamics is governed by a scalar potential $U(\phi^{j})$. In this case, the action is given by
\begin{eqnarray}
S = \int d^4x\sqrt{-g}\,\bigg[-\frac{1}{2}\,M_p^2\, R + \frac{1}{2}\,g^{\mu\nu}h_{ij}\partial_\mu \phi^{i}\,\partial_\nu \phi^{j} - U\bigg],\label{Ac}
\end{eqnarray}
where $h_{ij}(\phi^k)$ is the field space metric, $M_{p}$ is the 4d Planck mass, and the $4D$ space-time indexes $(\mu,\,\nu)$ are raised and lowered with the $4D$ metric $g_{\mu\nu}$ with a signature $(+,-,-,-)$, $U$ is a potential of an canonical normalized field. Very recently, the “refined” version of the swampland conjecture has been suggested \cite{Garg:2018reu,Ooguri:2018wrx}. It was found that the refined conjecture imposes a slightly weaker criterion on the scalar field potential
in inflation, and is consistent with the existence of a tachyonic instability. In light of the refined swampland conjecture, a scalar field potential associated with a self-consistent
UV-complete effective field theory must satisfy one of
the two conditions:
\begin{eqnarray}
|\nabla U|\geq \frac{c_{1}}{M_{p}}U\,,\label{R1}
\end{eqnarray}
or
\begin{eqnarray}
{\rm min}(\nabla_{j}\nabla_{j} U)\,\,\leq -\frac{c_{2}}{M^{2}_{p}}U\,,\label{R2}
\end{eqnarray}
where $c_1$ and $c_2$ are both positive constants of the order of ${\cal O}(1)$ and $|\nabla V|=\sqrt{g^{ij}\nabla_{j}\nabla_{j} V}$. The first condition corresponds to the original “swampland conjecture” proposed in Ref.\cite{Obied:2018sgi}. The conjectures require the above inequalities to be satisfied by any EFT which has a self-consistent ultraviolet (UV) completion. Notice that single-field inflation is apparently satisfied by the first condition, and the field variation is in excellent agreement with the well-known Lyth Bound for single-field inflation \cite{Lyth:1996im}. However, the second condition poses difficulties for single-field
paradigm \cite{Agrawal:2018own,Garg:2018reu,Dimopoulos:2018upl,Matsui:2018bsy}. This tension has been explored in
a number of recently related works, see for example Refs.\cite{Kehagias:2018uem,Achucarro:2018vey,Kinney:2018nny,Das:2018hqy,Ashoorioon:2018sqb,Brahma:2018hrd,Lin:2018rnx,Motaharfar:2018zyb,Lin:2018kjm,Holman:2018inr}. Interestingly, however, more complex models, e.g., multifield inflation \cite{Achucarro:2018vey} and warm inflation \cite{Motaharfar:2018zyb}, are still
allowed. Additionally, swampland criteria was also investigated of alternative theories of gravity, see for instance \cite{Yi:2018dhl,Heisenberg:2019qxz,Brahma:2019kch}. In \cite{Artymowski:2019vfy}, the viability of $f(R)$ and Brans-Dicke theories of gravity was also discussed. For any $U$, the standard slow-roll parameters can be rewritten using the inequalities to yield
\begin{eqnarray}
\sqrt{2\epsilon_{U}}\geq c_{1}\,,\quad{\rm or}\quad \eta_{U}\,\,\leq -c_{2}\,.\label{R12}
\end{eqnarray}
In the present work, we consider the implications of this slightly weaker constraint on a deformation of Starobinsky gravity. It is worth noting that a different model of deformed Starobinsky inflation was so far studied in Ref.\cite{Sebastiani:2013eqa}, while log corrections to $R^{2}$ gravity were investigated in Ref.\cite{Myrzakulov:2014hca,Bamba:2014jia}. Regarding the refined version, there exist relevant discussions of
general constraints on inflation and other cosmological/astrophysical models, see Refs.\cite{Kinney:2018kew,Haque:2019prw,Andriot:2018mav,Fukuda:2018haz,Garg:2018zdg,
Park:2018fuj,Schimmrigk:2018gch,Cheong:2018udx,
Chiang:2018lqx}.

This paper is organized as follows. In Sec.\ref{sec2}, we present
a summary of deformed Starobinsky gravity by following the work proposed by Ref.\cite{Codello:2014sua}. Here we linearize the action by introducing an auxiliary field method. We use the conformal transformations in order to transform the theory in the Jordan frame to the Einstein frame. The above transformation connects both theories and allows us to rewrite the action in terms of a  propagating  scalar  field  minimally  coupled  to  gravity. Sec.\ref{swa} is devoted to  discuss  the  theoretical  viability  of  deformed  Starobinky  gravity in light of the refined Swampland conjectures. We further discuss the refined Swampland criteria in deformed Starobinsky gravity in Sec.\ref{ref}. Our conclusions are reported in the last section.

%%%%%%%%%%%%%%%%%%%%%%%%%%%%%
\section{Deformed Starobinsky gravity: a short recap}\label{sec2}
%%%%%%%%%%%%%%%%%%%%%%%%%%%%%
Among many others, an intriguing possibility is that the gravity itself can be directly responsible for the inflationary period of the universe. The examination requires us to go beyond time-honored Einstein-Hilbert (EH) action. This can be achieved by adding a $R^{2}$ term to the original EH theory as in the Starobinsky model \cite{Starobinsky:1980te}. The model is highly natural since gravity itself drives cosmic inflation without the need of the scalar fields. It is worth noting that the model predicts a nearly vanishing ratio of tensor to scalar modes which is in excellent agreement with the observations, e.g. PLANCK data \cite{Ade:2015lrj,Akrami:2018odb}. Moreover, the logarithm corrections to $R^2$ gravity of the form $M^{2}_{p}R/2+(a/2)R^{2}/[1+b\ln(R/\mu^{2})]$, where $R$ is the Ricci scalar, $a$ and $b$ are constants and $\mu$ is an energy scale, suggested by asymptotic safety were recently considered in Ref.\cite{Liu:2018hno}.

In light of the observations, a discovery of primordial tensor modes can be used to constrain the cosmological parameters at the inflationary scale, which turn out to be close or at the grand unification energy scale. In general, the effective action for gravity can be in principle derived by considering the Taylor expansion in the Ricci scalar, $R$. Here without assuming a concrete form for the function $f(R)$, we consider
\ba
{\cal S}=\int d^{4}x\sqrt{-g}f(R)\equiv\int d^{4}x\sqrt{-g}\Big[a_{0}+a_{1}R+a_{2}R^{2}+...\Big]\,.\label{R}
\ea
The first term $a_{0}$ is like the cosmological constant and must be small. The next coefficient $a_{1}$ can be set to one as in general relativity. Regarding the Starobinsky gravity, we have $a_{2}=1/(6M^{2})$ where a constant $M$ has the dimensions of mass, see cosmological implications of the model \cite{Chatrabhuti:2015mws}. Here the ellipses may include the Weyl tensor $C^{2}$ and the Euler topological terms $E$. As mentioned in Ref.\cite{Codello:2014sua}, the $E$ terms can be ignored since it is just a total derivative. Moreover, the Weyl terms are subleading when gravity is quantized around a flat background. Higher powers of $R,\,C^{2}$ and $E$ are naturally suppressed by the Planck mass. Interestingly, the authors of Ref.\cite{Codello:2014sua} also take into account marginal deformations of the action (\ref{R}) by including logarithmic corrections. The authors consider a simple form of the gravitational action formulated in the Jordan frame:
\ba
{\cal S}_{J}=\int d^{4}x\sqrt{-g}\Big[-\frac{M^{2}_{p}}{2}R+hM^{4\alpha}_{p}R^{2(1-\alpha)}\Big]\,,\label{RJ}
\ea
where $h$ is a dimensionless parameter and $\alpha$ is a real parameter which is assumed as $2|\alpha|<1$. Note that the condition of the parameter $\alpha$ is further examined in the context of gravity's rainbow \cite{Channuie:2019kus}. One can linearize the above action by introducing an auxiliary field $y$ such that ${\cal S}_{J}=\int d^{4}x\sqrt{-g}\Big[f(y)+f'(y)(R-y)\Big]$ with $f(R)=-M^{2}_{p}R/2+hM^{4\alpha}_{p}R^{2(1-\alpha)}$ where $f'(y)=df(y)/dy$. Here the equation of motion for $y$ implies $R=y$ provided $f''(y)$ does not vanish. The explicit relation between (\ref{R}) and the effective quantum-corrected nonminimally coupled scalar field theory used in Ref.\cite{Joergensen:2014rya} can be done by introducing the conformal mode $\psi=-f'(y)$ with $V(\psi)=-y(\psi)\psi-f(y(\psi))$ and having introduced the mass-dimension one real scalar field $\varphi$ via $2\psi-M^{2}_{p}=\xi\varphi^{2}$ \cite{Codello:2014sua} to obtain:
\ba
{\cal S}_{J}=\int d^{4}x\sqrt{-g}\Big[-\frac{M^{2}_{p}+\xi\varphi^{2}}{2}R+V(\varphi)\Big]\,,\label{RJp}
\ea
where
\ba
V(\varphi)=\lambda\varphi^{4}\Big(\frac{\varphi}{M_{p}}\Big)^{4\gamma}\,\,\,{\rm with}\,\,\,\alpha=\frac{\gamma}{1+2\gamma},\label{RJpa}
\ea
and
\ba
h^{1+2\gamma}=\Big(\frac{\xi(1+2\gamma)}{4(1+\gamma)}\Big)^{2(1+\gamma)}\frac{1}{\lambda(1+2\gamma)}.\label{RJpa}
\ea
Notice from Eq.(\ref{RJp}) that the kinetic term for the field $\varphi$ is absent in the Jordan frame. However introducing the following conformal transformation of the metric, the kinetic term of the field can be simply generated via:
\ba
{\tilde g}_{\mu\nu}=\Omega(\varphi)^{2}g_{\mu\nu}\,,\,\,{\rm with}\,\,\Omega^{2}=1+\frac{\xi\varphi^{2}}{M^{2}_{p}}.\label{RJpa}
\ea
The above transformation connects both theories and allows us to rewrite the action in terms of a propagating scalar field minimally coupled to gravity. The resulting action is written in the Einstein frame and takes the form:
\ba
{\cal S}_{E}=\int d^{4}x\sqrt{-g}\Big[-\frac{M^{2}_{p}}{2}R+\frac{1}{2}g^{\mu\nu}\partial_{\mu}\chi\partial_{\nu}\chi-U(\chi)\Big]\,,\,\,\,U(\chi)=\Omega^{-4}V(\varphi(\chi)).\label{REp}
\ea
Notice that the action is written in terms of the canonically normalized field $\chi$ which is related to $\varphi$ via \cite{Codello:2014sua}
\ba
\frac{1}{2}\Big(\frac{d\chi}{d\varphi}\Big)^{2}=\frac{M^{2}_{p}(\sigma M^{2}_{p}+(\sigma+3\xi)\xi\varphi^{2})}{(M^{2}_{p}+\xi\varphi^{2})^{2}}.\label{REp}
\ea
It is worth noting that when setting $\sigma = 0$ a map from the Jordan frame of $f(R)$ gravity
to the Einstein frame with a canonically normalized field is obtained. Throughout this work, we will set $\sigma=0$. An explicit
relation between $\chi$ and $\varphi$ can be obtained by assuming that inflation occurs at large values of the scalar field, i.e. $\varphi\gg M_{p}/\sqrt{\xi}$ and we obtain
\ba
\chi\simeq \sqrt{6} M_{p}\log\Big[\frac{\sqrt{\xi}\varphi}{M_{p}}\Big].
\ea
Substituting the above canonical normalized field into Eq.(\ref{RJpa}), therefore the Einstein frame potential
takes the form
\ba
U(\chi) = \frac{\lambda  M_p^4}{\xi^{2}} e^{\frac{2 \sqrt{\frac{2}{3}}\chi}{M_p}}\Big(e^{\frac{\sqrt{\frac{2}{3}} \chi }{M_p}}+1\Big)^{-2}\,e^{\frac{4\gamma\chi}{\sqrt{6} M_p}} \xi^{-2\gamma}\,,\,\,\gamma=\frac{\alpha}{1-2\alpha}.\label{UU}
\ea
It is worth noting that in the limit of $\gamma=0$ (or equivalently $\alpha=0$) one recovers the Starobinsky model \cite{Starobinsky:1980te}. It's pointed out in Ref.\cite{Codello:2014sua} that for $0 < \alpha < 0.5$ the potential grows
exponentially, and then an inflationary model with nonzero primordial tensor modes can be successfully obtained. Note that the exact Einstein-frame potential of $f(R)=R+\lambda R^{p}$ with $p$ being not necessarily an integer in general was also derived in Ref.\cite{Motohashi:2014tra}.

%%%%%%%%%%%%%%%%%%%%%%%%%%%%%%%%%%%
\section{The refined Swampland criteria in deformed Starobinsky gravity}\label{swa}
%%%%%%%%%%%%%%%%%%%%%%%%%%%%%%%%%%
In the present work, we test a model of inflation of deformed Starobinsky gravity in general scalar-tensor theories of gravity. For our analysis below, we first define two new parameters for any scalar field $U(\chi)$:
\begin{eqnarray}
F_{1}=\frac{|dU(\chi)/d\chi|}{U(\chi)}\,,
\label{f1}
\end{eqnarray}
and
\begin{eqnarray}
F_{2}=\frac{d^{2}U(\chi)/d\chi^{2}}{U(\chi)}\,.
\label{f2}
\end{eqnarray}
Considering Eq.(\ref{R12}), the above parameters can be recast in terms of the slow-roll parameters to yield
\begin{eqnarray}
F_{1}=\sqrt{2\varepsilon_{U}}\,,\quad F_{2}=\eta_{U}\,.
\label{f1f2}
\end{eqnarray}
We see that $F_1$ and $F_2$ are written in terms of the slow-roll parameters. Therefore, they can be related to the spectrum index of the primordial curvature power spectrum $n_{s}$ and tensor-to-scalar ratio $r$. In the present case, it is rather straightforward to show that
\begin{eqnarray}
F_{1}=\sqrt{2\varepsilon_{U}}=\sqrt{\frac{r}{8}}\,,
\label{f11}
\end{eqnarray}
and
\begin{eqnarray}
F_{2}=\eta_{U}=\frac{1}{2}\big(n_{s}-1+\frac{3}{8}r\big)\,.
\label{f12}
\end{eqnarray}
Below we examine if deformed Starobinsky inflation does satisfy this new refined swampland conjecture, or not. In establishing the connection among the swampland conditions and the parameters of the model considered, we consider the well-known inflationary parameters, i.e., the scalar spectral
index $n_{s}$, and tensor to scalar ratio $r$, and from the standard formulation we write \ba
n_{s}=1-6\epsilon_{U}+2\eta_{U},\,\,r=16\epsilon_{U}, \label{ns}
\ea
where in terms of the potential the slow-roll parameters can be defined as
\ba
\epsilon_{U}=\frac{M^{2}_{p}}{2}\left(\frac{U'}{U}\right)^{2}\,,\,\,\eta_{U}=M^{2}_{p}\left(\frac{U''}{U}\right),\label{epe}
\ea
where primes denote derivatives with respect to the field $\chi$. It is useful to express $\varepsilon_{U}$ and $\eta_{U}$ in terms of the number of e-foldings. Substituting Eq.(\ref{UU})
into Eq.(\ref{epe}), we find up to the first order of $\gamma$:
\ba
\epsilon_{U}&=& \frac{3}{4 N^2}+\frac{\gamma }{N}+{\cal O}(\gamma^{2}),\\\eta_{U}&= &-\frac{1}{N}+\frac{2 \gamma }{3}+{\cal O}(\gamma^{2})\,.\label{epeN}
\ea
We notice that when setting $\gamma=0$ the results recover Starobinsky inflation. Using the above expressions, we can rewrite Eq.(\ref{ns}) in terms of the the number of e-foldings, $N$, as
\ba
n_{s}&\simeq&1-6\Big(\frac{3}{4 N^2}+\frac{\gamma }{N}\Big)+2\Big(-\frac{1}{N}+\frac{2 \gamma }{3}\Big),\\r&\simeq& 16\Big(\frac{3}{4 N^2}+\frac{\gamma }{N}\Big),\label{nsrN}
\ea
where we have used the potential from Eq.(\ref{UU}). {\color{blue}{Note here that the very small tensor-to-scalar ratio $r=12\,N^2$ for the Starobinsky $R+R^2$ inflationary model was first (and quantitatively correctly) presented in Ref.\cite{Starobinsky:1983zz}}}. Interestingly, usual $\phi^{4}$ inflation refers to the results when setting $\gamma=0$, that is, non-minimally
coupled $\phi$ inflation. Here a term with $\gamma$ is used to clarify how the results deviate from $\phi^{4}$ inflation. An expansion is, however, justified for tiny values of $\gamma$. We then constrain values of $\gamma$ using a condition of $r<0.06$
\begin{eqnarray}
r\simeq16\Big(\frac{3}{4 N^2}+\frac{\gamma }{N}\Big)\,<\,0.06\,,
\end{eqnarray}
to obtain
\begin{eqnarray}
\gamma <\frac{0.002 \left(2 N^2-375\right)}{N}\quad {\rm or}\quad \alpha < \frac{2\,N^2-375}{4\,N^2+500\,N-750}.
\end{eqnarray}
Assuming $N=55$ and $\alpha=0.1461$, this model predicts $n_{s}\simeq 0.965$ and $r\simeq 0.00464$ which are consistent with the observed data \cite{Planck:2018jri}. Inserting these values into Eq.(\ref{f1}) and Eq.(\ref{f2}), we obtain
\begin{eqnarray}
F_{1}&=&\sqrt{2\varepsilon_{U}}=\sqrt{\frac{r}{8}}=0.02410\,,\\F_{2}&=&\eta_{U}=\frac{1}{2}\big(n_{s}-1+3r/8\big)=-0.02596\,.
\label{f1f2}
\end{eqnarray}
Considering the refined swanpland conjecture (\ref{ReRe}), we find
\begin{eqnarray}
c_{1}\leq 0.02410\quad{\rm or}\quad c_{2}\leq 0.02596\,.
\label{c1c23}
\end{eqnarray}
Clesarly, neither $c_{1}$ nor $c_{2}$ are of the order of ${\cal O}(1)$ implying that the model under consideration is in strong tension with the refined swampland conjecture.

%%%%%%%%%%%%%%%%%%%%%%%%
\section{Refining de Sitter swampland conjecture}\label{ref}
%%%%%%%%%%%%%%%%%%%%%%%%
Very recently, the authors of \cite{Andriot:2018mav} have proposed a single condition on both $\epsilon_{U}$ and $\eta_{U}$ has been proposed. The approach is called a further refining de Sitter swampland conjecture. The statement of an alternative refined de Sitter conjecture is suggested that a low energy effective theory of a quantum gravity that takes the form (\ref{REp}) should verify, at any point in field space where $U>0$ \cite{Andriot:2018mav},
\begin{eqnarray}
\Big(M_{p}\frac{|\nabla U|}{U}\Big)^{q}-a\,M^{2}_{p}\frac{{\rm min}(\nabla_{j}\nabla_{j} U)}{U}\geq b\,\quad{\rm with}\quad a+b=1,\,a,\,b>0,\,q>2\,,\label{ReRe}
\end{eqnarray}
in which a combination of the first and second derivatives of the scalar potential is achieved. In terms of the slow-roll parameters, the conjecture implies \cite{Andriot:2018mav}
\begin{eqnarray}
(2\epsilon_{U})^{q/2}-a\,\eta_{U}\geq b\,.\label{Re1}
\end{eqnarray}
Interestingly, the authors of Ref.\cite{Liu:2021diz} have discovered that Higgs inflation model, Palatini Higgs inflation, and Higgs-Dilaton model can always satisfy this new swampland conjecture if only they adjust the relevant parameters $a,\,b=1-a$ and $q$. Substituting the above results into Eq.(\ref{Re1}), we have
\begin{eqnarray}
0.02408^{q}+0.01663\,a\geq 1-a\,\quad{\rm or}\quad 0.02408^{q}\geq 1-1.01663\,a\,.
\label{re111}
\end{eqnarray}
If we can find $a$ to satisfy the condition
\begin{eqnarray}
\frac{1}{1.01663}(1-0.02408^{q})\leq a<1\,,
\quad q>2,
\label{r111}
\end{eqnarray}
then the further refining swampland conjecture can be satisfied. In this case, when $a=\nicefrac{1}{1.01663}$, we have $1-1.01663\,a=0$. Therefore, we can examine that when $a < \nicefrac{1}{1.01663}$, we can always find a $q$ whose value is larger than $2$. It is possible to give an example of values of the parameters $a,\,b,\,q$, which work for this model. From Eq.(\ref{r111}), we use $q=2.2$ which is satisfied by a condition $q>2$. We find for this particular case that $0.983371 \leq a<1$ and choose $a=0.9834<\nicefrac{1}{1.01663}=0.98364$ and $1-a=1-0.9834=b=0.0166>0$.

\section{Conclusion}
Recently, the Swampland conjecture has attracted significant attention. The conjecture allows us to validate or invalidate a large class of low energy effective theories. It can be formulated as inequalities on the potential of a scalar field which is satisfied by certain constraints. In this work, We first considered the deformation of the form $f(R)\sim R^{2(1-\alpha)}$ with $\alpha$ being a constant. We then constrained a parameter $\alpha$ using the spectral index of curvature perturbation $n_s$ and the tensor-to-scalar ratio $r$. Here we linearized the original action by introducing an auxiliary field method and used the conformal transformations in order to transform the theory in the Jordan frame to the Einstein frame. We rewrote the action in terms of the cannonical normalized  scalar field  minimally  coupled  to  gravity.

We discussed the theoretical viability of deformed Starobinky gravity in light of the recent refined Swampland conjectures. Our analysis showed that the model under consideration is in strong tension with the refined swampland conjecture. However, regarding our analysis with proper choices of parameters  $a,\,b=1- a$ and $q$, we discovered that the model can always satisfy this new refined swampland conjecture. Therefore, the model might be in “landscape” since the “further refining de Sitter swampland conjecture” is satisfied.

\acknowledgments
This research was partially supported by the New Strategic Research (P2P) project, Walailak
University, Thailand.

%%%%%%%%%%%%%%%%%%%%%%%%%%%%%%%%%%%%%%%%
%%%%%%%%%%%%%%%%%%%%%%%%%%%%%%%%%%%%%%%%

\begin{thebibliography}{99}

%\cite{Obied:2018sgi}
\bibitem{Obied:2018sgi} 
  G.~Obied, H.~Ooguri, L.~Spodyneiko and C.~Vafa,
  %``De Sitter Space and the Swampland,''
  arXiv:1806.08362 [hep-th]
  
  %\cite{Agrawal:2018own}
\bibitem{Agrawal:2018own} 
  P.~Agrawal, G.~Obied, P.~J.~Steinhardt and C.~Vafa,
  %``On the Cosmological Implications of the String Swampland,''
  Phys.\ Lett.\ B {\bf 784}, 271 (2018)
  
    
  %\cite{Ooguri:2018wrx}
\bibitem{Ooguri:2018wrx} 
  H.~Ooguri, E.~Palti, G.~Shiu and C.~Vafa,
  %``Distance and de Sitter Conjectures on the Swampland,''
  Phys.\ Lett.\ B {\bf 788}, 180 (2019)

  %\cite{Palti:2019pca}
\bibitem{Palti:2019pca} 
  E.~Palti,
  %``The Swampland: Introduction and Review,''
  Fortsch.\ Phys.\  {\bf 67}, no. 6, 1900037 (2019)
  
  %\cite{Lyth:1996im}
\bibitem{Lyth:1996im} 
  D.~H.~Lyth,
  %``What would we learn by detecting a gravitational wave signal in the cosmic microwave background anisotropy?,''
  Phys.\ Rev.\ Lett.\  {\bf 78}, 1861 (1997)
  
%%%%%Single Field Inflation%%%%%
  
  %\cite{Garg:2018reu}
\bibitem{Garg:2018reu} 
  S.~K.~Garg and C.~Krishnan,
  %``Bounds on Slow Roll and the de Sitter Swampland,''
  arXiv:1807.05193 [hep-th]
  
  %\cite{Achucarro:2018vey}
\bibitem{Achucarro:2018vey} 
  A.~Achúcarro and G.~A.~Palma,
  %``The string swampland constraints require multi-field inflation,''
  JCAP {\bf 1902}, 041 (2019)
  
  %\cite{Motaharfar:2018zyb}
\bibitem{Motaharfar:2018zyb} 
  M.~Motaharfar, V.~Kamali and R.~O.~Ramos,
  %``Warm inflation as a way out of the swampland,''
  Phys.\ Rev.\ D {\bf 99}, no. 6, 063513 (2019)


%%%Others model
%\cite{Kehagias:2018uem}
\bibitem{Kehagias:2018uem} 
  A.~Kehagias and A.~Riotto,
  %``A note on Inflation and the Swampland,''
  Fortsch.\ Phys.\  {\bf 66}, no. 10, 1800052 (2018)
  
  %\cite{Achucarro:2018vey}
\bibitem{Achucarro:2018vey} 
  A.~Achúcarro and G.~A.~Palma,
  %``The string swampland constraints require multi-field inflation,''
  JCAP {\bf 1902}, 041 (2019)
  
  %\cite{Kinney:2018nny}
\bibitem{Kinney:2018nny} 
  W.~H.~Kinney, S.~Vagnozzi and L.~Visinelli,
  %``The zoo plot meets the swampland: mutual (in)consistency of single-field inflation, string conjectures, and cosmological data,''
  Class.\ Quant.\ Grav.\  {\bf 36}, no. 11, 117001 (2019)
  
  %\cite{Das:2018hqy}
\bibitem{Das:2018hqy} 
  S.~Das,
  %``Note on single-field inflation and the swampland criteria,''
  Phys.\ Rev.\ D {\bf 99}, no. 8, 083510 (2019)
  
  %\cite{Ashoorioon:2018sqb}
\bibitem{Ashoorioon:2018sqb} 
  A.~Ashoorioon,
  %``Rescuing Single Field Inflation from the Swampland,''
  Phys.\ Lett.\ B {\bf 790}, 568 (2019)
  
  %\cite{Brahma:2018hrd}
\bibitem{Brahma:2018hrd} 
  S.~Brahma and M.~Wali Hossain,
  %``Avoiding the string swampland in single-field inflation: Excited initial states,''
  JHEP {\bf 1903}, 006 (2019)
  
  %\cite{Lin:2018rnx}
\bibitem{Lin:2018rnx} 
  C.~M.~Lin,
  %``Type I Hilltop Inflation and the Refined Swampland Criteria,''
  Phys.\ Rev.\ D {\bf 99}, no. 2, 023519 (2019)
  
  %\cite{Motaharfar:2018zyb}
\bibitem{Motaharfar:2018zyb} 
  M.~Motaharfar, V.~Kamali and R.~O.~Ramos,
  %``Warm inflation as a way out of the swampland,''
  Phys.\ Rev.\ D {\bf 99}, no. 6, 063513 (2019)
  
  %\cite{Lin:2018kjm}
\bibitem{Lin:2018kjm} 
  C.~M.~Lin, K.~W.~Ng and K.~Cheung,
  %``Chaotic inflation on the brane and the Swampland Criteria,''
  arXiv:1810.01644 [hep-ph]
  
  %\cite{Holman:2018inr}
\bibitem{Holman:2018inr} 
  R.~Holman and B.~Richard,
  %``Spinodal solution to swampland inflationary constraints,''
  Phys.\ Rev.\ D {\bf 99}, no. 10, 103508 (2019)

%%%% others modified gravity

  %\cite{Yi:2018dhl}
\bibitem{Yi:2018dhl} 
  Z.~Yi and Y.~Gong,
  %``Gauss-Bonnet inflation and swampland,''
  arXiv:1811.01625 [gr-qc]

%\cite{Heisenberg:2019qxz}
\bibitem{Heisenberg:2019qxz} 
  L.~Heisenberg, M.~Bartelmann, R.~Brandenberger and A.~Refregier,
  %``Horndeski gravity in the swampland,''
  Phys.\ Rev.\ D {\bf 99}, no. 12, 124020 (2019)
  
  %\cite{Brahma:2019kch}
\bibitem{Brahma:2019kch} 
  S.~Brahma and M.~W.~Hossain,
  %``Dark energy beyond quintessence: Constraints from the swampland,''
  JHEP {\bf 1906}, 070 (2019)
  
  %\cite{Artymowski:2019vfy}
\bibitem{Artymowski:2019vfy} 
  M.~Artymowski and I.~Ben-Dayan,
  %``f(R) and Brans-Dicke Theories and the Swampland,''
  JCAP {\bf 1905}, no. 05, 042 (2019)
  
   %new
  
  %\cite{Sebastiani:2013eqa}
\bibitem{Sebastiani:2013eqa} 
  L.~Sebastiani, G.~Cognola, R.~Myrzakulov, S.~D.~Odintsov and S.~Zerbini,
  %``Nearly Starobinsky inflation from modified gravity,''
  Phys.\ Rev.\ D {\bf 89}, no. 2, 023518 (2014)
  
  %\cite{Myrzakulov:2014hca}
\bibitem{Myrzakulov:2014hca} 
  R.~Myrzakulov, S.~Odintsov and L.~Sebastiani,
  %``Inflationary universe from higher-derivative quantum gravity,''
  Phys.\ Rev.\ D {\bf 91}, no. 8, 083529 (2015)
  
 % \cite{Bamba:2014jia}
\bibitem{Bamba:2014jia} 
  K.~Bamba, R.~Myrzakulov, S.~D.~Odintsov and L.~Sebastiani,
  %``Trace-anomaly driven inflation in modified gravity and the BICEP2 result,''
  Phys.\ Rev.\ D {\bf 90}, no. 4, 043505 (2014)
  
%%%%%%Others refined version

%\cite{Kinney:2018kew}
\bibitem{Kinney:2018kew} 
  W.~H.~Kinney,
  %``Eternal Inflation and the Refined Swampland Conjecture,''
  Phys.\ Rev.\ Lett.\  {\bf 122}, no. 8, 081302 (2019)
  
  %\cite{Haque:2019prw}
\bibitem{Haque:2019prw} 
  M.~R.~Haque and D.~Maity,
  %``Reheating constraints on the inflaton and dark matter: Swampland conjecture,''
  Phys.\ Rev.\ D {\bf 99}, no. 10, 103534 (2019)
  
%\cite{Andriot:2018mav}
\bibitem{Andriot:2018mav} 
  D.~Andriot and C.~Roupec,
  %``Further refining the de Sitter swampland conjecture,''
  Fortsch.\ Phys.\  {\bf 67}, no. 1-2, 1800105 (2019)
  
  %\cite{Fukuda:2018haz}
\bibitem{Fukuda:2018haz} 
  H.~Fukuda, R.~Saito, S.~Shirai and M.~Yamazaki,
  %``Phenomenological Consequences of the Refined Swampland Conjecture,''
  Phys.\ Rev.\ D {\bf 99}, no. 8, 083520 (2019)
  
  %\cite{Garg:2018zdg}
\bibitem{Garg:2018zdg} 
  S.~K.~Garg, C.~Krishnan and M.~Zaid Zaz,
  %``Bounds on Slow Roll at the Boundary of the Landscape,''
  JHEP {\bf 1903}, 029 (2019)
  
  %\cite{Park:2018fuj}
\bibitem{Park:2018fuj} 
  S.~C.~Park,
  %``Minimal gauge inflation and the refined Swampland conjecture,''
  JCAP {\bf 1901}, no. 01, 053 (2019)
  
  %\cite{Schimmrigk:2018gch}
\bibitem{Schimmrigk:2018gch} 
  R.~Schimmrigk,
  %``The Swampland Spectrum Conjecture in Inflation,''
  arXiv:1810.11699 [hep-th]
  
  %\cite{Cheong:2018udx}
\bibitem{Cheong:2018udx} 
  D.~Y.~Cheong, S.~M.~Lee and S.~C.~Park,
  %``Higgs Inflation and the Refined dS Conjecture,''
  Phys.\ Lett.\ B {\bf 789}, 336 (2019)
  
  %\cite{Chiang:2018lqx}
\bibitem{Chiang:2018lqx} 
  C.~I.~Chiang, J.~M.~Leedom and H.~Murayama,
  %``What does Inflation say about Dark Energy given the Swampland Conjectures?,''
  arXiv:1811.01987 [hep-th]

%%%%%%%%%Starobinsky%%%%%%%%%%%
%\cite{Starobinsky:1980te}
\bibitem{Starobinsky:1980te} 
  A.~A.~Starobinsky,
  %``A New Type of Isotropic Cosmological Models Without Singularity,''
  Phys.\ Lett.\ B {\bf 91}, 99 (1980)
  [Phys.\ Lett.\  {\bf 91B}, 99 (1980)]
  [Adv.\ Ser.\ Astrophys.\ Cosmol.\  {\bf 3}, 130 (1987)]
  
  %\cite{Ade:2015lrj}
\bibitem{Ade:2015lrj} 
  P.~A.~R.~Ade {\it et al.} [Planck Collaboration],
  %``Planck 2015 results. XX. Constraints on inflation,''
  Astron.\ Astrophys.\  {\bf 594}, A20 (2016)
  
  %\cite{Akrami:2018odb}
\bibitem{Akrami:2018odb} 
  Y.~Akrami {\it et al.} [Planck Collaboration],
  %``Planck 2018 results. X. Constraints on inflation,''
  arXiv:1807.06211 [astro-ph.CO]
  
  %\cite{Liu:2018hno}
\bibitem{Liu:2018hno}
L.~H.~Liu, T.~Prokopec and A.~A.~Starobinsky,
%``Inflation in an effective gravitational model and asymptotic safety,''
Phys. Rev. D \textbf{98} (2018) no.4, 043505
%doi:10.1103/PhysRevD.98.043505
[arXiv:1806.05407 [gr-qc]].
%32 citations counted in INSPIRE as of 12 May 2022
  
  %%%%%%%%%%StaroRain
  
  %\cite{Chatrabhuti:2015mws}
\bibitem{Chatrabhuti:2015mws} 
  A.~Chatrabhuti, V.~Yingcharoenrat and P.~Channuie,
  %``Starobinsky Model in Rainbow Gravity,''
  Phys.\ Rev.\ D {\bf 93}, no. 4, 043515 (2016)
  
  %\cite{Codello:2014sua}
\bibitem{Codello:2014sua}
  A.~Codello, J.~Joergensen, F.~Sannino and O.~Svendsen,
  %``Marginally Deformed Starobinsky Gravity,''
  JHEP {\bf 1502} (2015) 050
  
  %\cite{Channuie:2019kus}
\bibitem{Channuie:2019kus} 
  P.~Channuie,
  %``Deformed Starobinsky model in gravity’s rainbow,''
  Eur.\ Phys.\ J.\ C {\bf 79}, no. 6, 508 (2019)

%\cite{Joergensen:2014rya}
\bibitem{Joergensen:2014rya} 
  J.~Joergensen, F.~Sannino and O.~Svendsen,
  %``Primordial tensor modes from quantum corrected inflation,''
  Phys.\ Rev.\ D {\bf 90}, no. 4, 043509 (2014)
  
  
  %\cite{Motohashi:2014tra}
\bibitem{Motohashi:2014tra} 
  H.~Motohashi,
  %``Consistency relation for $R^p$ inflation,''
  Phys.\ Rev.\ D {\bf 91}, 064016 (2015)
  
  %\cite{Starobinsky:1983zz}
\bibitem{Starobinsky:1983zz}
A.~A.~Starobinsky,
%``The Perturbation Spectrum Evolving from a Nonsingular Initially De-Sitter Cosmology and the Microwave Background Anisotropy,''
Sov. Astron. Lett. \textbf{9} (1983), 302
%425 citations counted in INSPIRE as of 12 May 2022

%%%%%%%%%%%%%%%%
  
  %\cite{Matsui:2018bsy}
\bibitem{Matsui:2018bsy} 
  H.~Matsui and F.~Takahashi,
  %``Eternal Inflation and Swampland Conjectures,''
  Phys.\ Rev.\ D {\bf 99}, no. 2, 023533 (2019)
  
  %\cite{Dimopoulos:2018upl}
\bibitem{Dimopoulos:2018upl} 
  K.~Dimopoulos,
  %``Steep Eternal Inflation and the Swampland,''
  Phys.\ Rev.\ D {\bf 98}, no. 12, 123516 (2018)
  
  %\cite{Andriot:2018mav}
\bibitem{Andriot:2018mav}
D.~Andriot and C.~Roupec,
%``Further refining the de Sitter swampland conjecture,''
Fortsch. Phys. \textbf{67} (2019) no.1-2, 1800105
%doi:10.1002/prop.201800105
[arXiv:1811.08889 [hep-th]].
%78 citations counted in INSPIRE as of 05 Feb 2022

%\cite{Liu:2021diz}
\bibitem{Liu:2021diz}
Y.~Liu,
%``Higgs inflation and its extensions and the further refining dS swampland conjecture,''
Eur. Phys. J. C \textbf{81} (2021) no.12, 1122
%doi:10.1140/epjc/s10052-021-09940-w
[arXiv:2112.14571 [hep-th]].
%0 citations counted in INSPIRE as of 05 Feb 2022

%\cite{Planck:2018jri}
\bibitem{Planck:2018jri}
Y.~Akrami \textit{et al.} [Planck],
%``Planck 2018 results. X. Constraints on inflation,''
Astron. Astrophys. \textbf{641} (2020), A10
%doi:10.1051/0004-6361/201833887
[arXiv:1807.06211 [astro-ph.CO]].
%1691 citations counted in INSPIRE as of 05 Feb 2022
  
\end{thebibliography}
\end{document}